\documentclass[pra,twocolumn,superscriptaddress,longbibliography]{revtex4-1}
\usepackage{amssymb,amsmath,amsfonts,bm}
\usepackage{epsfig,graphicx}
\usepackage{times}

\begin{document}

\title{Nonadiabatic Holonomic Quantum Computation with Dressed-State Qubits}

\author{Zheng-Yuan Xue} \email{zyxue@scnu.edu.cn}

\author{Feng-Lei Gu}

\author{Zhuo-Ping Hong}
\affiliation{Guangdong Provincial Key Laboratory of Quantum Engineering and Quantum Materials, and School of Physics\\ and Telecommunication Engineering, South China Normal University, Guangzhou 510006, China}

\author{Zi-He Yang}
\affiliation{School of Physics, Huazhong University of Science and Technology, Wuhan 430074, China}

\author{Dan-Wei Zhang} 
\affiliation{Guangdong Provincial Key Laboratory of Quantum Engineering and Quantum Materials, and School of Physics\\ and Telecommunication Engineering, South China Normal University, Guangzhou 510006, China}

\author{Yong Hu} \email{huyong@mail.hust.edu.cn}
\affiliation{School of Physics, Huazhong University of Science and Technology, Wuhan 430074, China}

\author{J. Q. You} \email{jqyou@csrc.ac.cn}
\affiliation{Quantum Physics and Quantum Information Division, Beijing Computational Science Research Center, Beijing 100193, China}

\date{\today}

\begin{abstract}
Implementing holonomic quantum computation is a challenging task as it requires complicated interaction among multilevel systems. Here we propose to implement nonadiabatic holonomic quantum computation based on dressed-state qubits in circuit QED. An arbitrary holonomic single-qubit gate can be conveniently achieved using external microwave fields and tuning their amplitudes and phases. Meanwhile, nontrivial two-qubit gates can be implemented in a coupled-cavities scenario assisted by a grounding superconducting quantum-interference device (SQUID) with tunable interaction, where the tuning is achieved by modulating the ac flux threaded through the SQUID. In addition, our proposal is directly scalable, up to a two-dimensional lattice configuration. In the present scheme,  the dressed states involve only the lowest two levels of each transmon qubit and  the effective interactions exploited are all of resonant nature. Therefore, we release the main difficulties for physical implementation of holonomic quantum computation on superconducting circuits.
\end{abstract}


\maketitle

\section{Introduction}
The superconducting quantum circuit (SQC) \cite{MakhlinRMP2001, ClarkeReviewNature2008, JQYouNatureReview2011, DevoretScienceReview2013} is a promising candidate for physical implementation of quantum computation due to its flexibility and scalability. However, the noises from the environment severely hinder the performance of quantum gates. On the other hand, geometric phase and holonomy depend only on the global property of the evolution trajectory and, thus, are insensitive to certain types of control errors \cite{ps2004,ps2012,noise2012, LeibfriedGeometricNature2003,DuJFGeometricPRA2006, SchoelkopfScience2007,WallraffHOGeomericPRL2012,GasparinettieSA2016}. This insensitivity  is one of the main advantages when implementing quantum computation in a geometric way, as the control lines and devices in a large-scale lattice will inevitably induce local noises and reduce the fidelity of dynamical quantum-gate operations. Therefore, holonomic quantum computation (HQC)  \cite{ZanardiHQC1999PLA,ZanardiHQCPRA1999, JonesGeometricNature2000, DuanLMScience2001, JiangLHQCPRL2016}, where quantum gates are induced by geometric transformations, has emerged as a potential way for robust quantum computing. To obtain an adiabatic geometric phase, it requires that the trajectory should travel under the adiabatic condition and, consequently, the required gate times are on the same order of the coherent times in typical physical systems \cite{WangXBNA2001PRL,ZhuWangNAGPRL2002}. Therefore, increasing research efforts have recently been devoted to nonadiabatic HQC \cite{SjoqvistNJP2012,XuGF2012PRL,mva2,ZhangJ2014PRA,LiangXue2014PRA, zhouj, xue1, WangYMPRA2016,xue2,es2016,zpz2016,mva1}, and some preliminary quantum gates were demonstrated in several  experiments \cite{FengLong2013PRL,AbdumalikovNature2013,ZuChongNature2014,sac,e7}. Nevertheless, due to the complicated interaction needed for implementing two-qubit gates, up to now only single-qubit holonomic gates have been experimentally demonstrated  on SQCs \cite{AbdumalikovNature2013}. Existing theoretical investigations of two-qubit holonomic gates usually use multilevel systems and result in a slow dispersive gate construction. This is particularly difficult for SQCs, as the anharmonicity of the energy spectrum of superconducting transmon qubits has been reduced to gain robustness against charge-type $1/f$ noises \cite{KochTransmonPRA2007,FlickerRMP2014}. This small anharmonicity limits the coupling strengths one can exploit and makes the implementation of universal HQC with SQC very inefficient.

Here, we present a practical scheme for nonadiabatic HQC in a circuit QED lattice, where we encode the logical qubits by dressed states built by transmission line resonators (TLRs) coupled with their transmons \cite{KochTransmonPRA2007}. In particular, the arbitrary single logical qubit operation can be obtained through the proper ac driving of the transmon qubit. More important, we propose the nontrivial two-qubit gate through the resonant interaction between TLRs of the logical qubits, which can be induced by a grounding superconducting quantum-interference device (SQUID) with a single-frequency ac magnetic modulation \cite{FelicettiPRL2014,WangYPChiral2015,WangYPNPJQI2015, WangYPLieb2016}. The distinct merit of our scheme is that it involves only the lowest two levels of the transmon qubits and can result in universal HQC in an all-resonant way, thus leading to fast and high-fidelity gates in a simple setup. Therefore, our proposal opens up the possibility of universal HQC on SQC, which can be immediately tested experimentally as it requires only the current state-of-art technology. The current proposal is essentially different from our previous  scheme \cite{xue2}, where the two-qubit gate is  implemented between the logical qubits defined by the decoherence-free subspace encoding. In addition, more ac modulations of the grounding SQUID are needed in Ref. \cite{xue2}, and the induced interactions of the logical qubits are complicated as well.

\section{The system and the logical qubit}
\begin{figure}[tb]
\centering
\includegraphics[width=7.5cm]{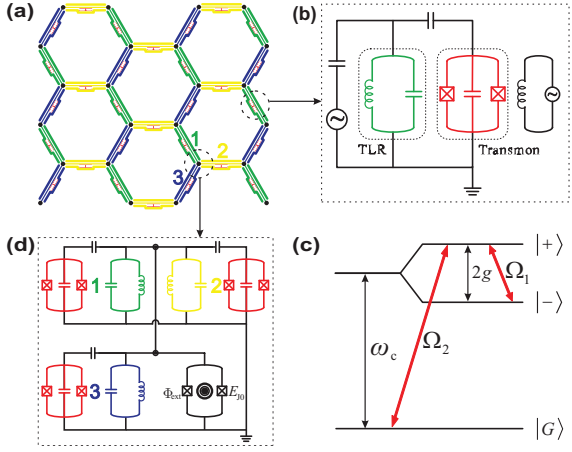}
\caption{Proposed circuit setup of scalable nonadiabatic HQC. (a) Two-dimensional
lattice consisting of three types of TLRs  placed in an interlaced honeycomb form. (b) Logic qubit built by the coupled TLR-transmon unit. (c) Energy level and driving configuration for the single-qubit gates in the dressed-state basis. (d) Coupling of the three TLRs at their common ends by a grounding SQUID, which is the building block of the 2D lattice. Through the modulation of the SQUID, the two-qubit gate between two dressed-state qubits can be realized.}
\label{Fig setup}
\end{figure}

We propose to realize the scalable HQC on a circuit QED lattice shown in Fig.~\ref{Fig setup}(a), which consists of three types of TLRs differed by their lengths and placed in an interlaced honeycomb form. At their ends, the TLRs are grounded by SQUIDs with effective inductances much smaller than those of the TLRs. The role of the grounding SQUIDs is to establish the well-separated TLR modes on this coupled lattice and to induce the consequent coupling between them \cite{FelicettiPRL2014,WangYPChiral2015,WangYPNPJQI2015,WangYPLieb2016}. We specify the eigenfrequencies of the three types of TLRs as $(\omega_{c1},\omega_ {c2},\omega_{c3})=(\omega_{c}, \omega_{c}+3\delta_c, \omega_{c}+\delta_c)$ with $\omega_{c}/2\pi = 6$  GHz and $\delta_c/2\pi = 0.4$ GHz. Such frequency configuration is for the following application of parametric coupling and can be experimentally realized through the length selection of the TLRs \cite{NISTParametricConversionNP2011,NISTHongOuMandelPRL2012, NISTParametricCouplingPRL2014,NISTCoherentStateAPL2015}.
In addition, we introduce for each TLR a transmon qubit with its eigenfrequency tunable through the modulation of its Josephson coupling energy and the TLR-transmon coupling strength which can reach the strong coupling region \cite{DevoretScienceReview2013}. The logical qubit of our scheme is physically formed by the basic building block of the lattice, i.e., each TLR together with its transmon, as shown in Fig. \ref{Fig setup}(b). Taking the particular TLR-transmon unit in Fig. \ref{Fig setup}(b) as an example, we can describe it by the Jaynes-Cummings Hamiltonian
\begin{eqnarray}
\label{Eqn JC}
H_{\text{JC}} ={1 \over 2} \hbar \omega_{q} \sigma_z
+ \hbar \omega_{c} a^{\dagger}a +g_0 (a\sigma^{+} + a^{\dagger} \sigma^{-}),
\end{eqnarray}
where $\omega_{q}$ is the eigenfrequency of the transmon qubit, $\sigma^{\pm}$ and $\sigma_{z}$ are the Pauli operators of the transmon qubit, $a^\dagger$ and $a$ are the creation and annihilation operators of the TLR, and $g_0$ is the transmon-TLR coupling strength. In the resonant condition $\omega_{q}=\omega_{c}$, the first three lowest eigenstates of the system are $|G\rangle=|0\rangle_{q}|0\rangle_{c}$ and ${| \pm \rangle}=\left(|0\rangle_{q}|1\rangle_{c} \pm |1\rangle_{q}|0\rangle_{c}\right)/\sqrt{2}$ with eigenenergies $E_{G}=0$ and  $E_\pm=\hbar \omega_{c} \pm g_0$, respectively [Fig. \ref{Fig setup}(c)]. Hereafter, we encode the logic qubit by span$\{ |G\rangle, | - \rangle \}$ and exploit $| + \rangle$ as an ancillary state.

\section{The Universal single-qubit gates}

The single-qubit nonadiabatic holonomic gates can be established through a two-tone microwave driving
\begin{eqnarray}
H_{d} = 2f_1(t) \sigma_z + 2\sqrt{2} f_2(t) \sigma_x,
\label{Eqn Hd}
\end{eqnarray}
on the transmon qubit, with $f_1(t)= \hbar \Omega_{1}\cos(2g_0 t)$, $f_2(t)=\hbar \Omega_{2}\cos(E_{+} t+\varphi)$, $\Omega_{1,2}$ being the amplitudes of the two tones, and $\varphi$ being a prescribed phase factor. The $\sigma_x$ tone connecting the $| G \rangle \Leftrightarrow | + \rangle$ transition can be induced by the capacitive link of the external ac pulses to the transmon qubit, and the $\sigma_z$ tone connecting the $| - \rangle \Leftrightarrow | + \rangle$ transition can be accomplished via the modulation of the Josephson energy of the transmon through its magnetic flux bias [Fig. \ref{Fig setup}(b)]. The reduced Hamiltonian in the subspace span $\{ |G\rangle, |-\rangle, |+\rangle \}$ takes the form of
\begin{eqnarray}\label{h1}
H_{1}&=&H_{\text{JC}}+H_{d}\\
&=& 2 \left(
  \begin{array}{ccc}
    0 &  -f_2(t) &  f_2(t)\\
    -f_2(t) & E_{-} & -f_1(t) \\
   f_2(t)& -f_1(t) & E_{+} \\
  \end{array}
\right). \notag
\end{eqnarray}
Assuming $g_0 \gg\Omega =\sqrt{\Omega_{1}^2+\Omega_{2}^2}$, we can obtain in the rotating frame of $H_\mathrm{JC}$ the effective Hamiltonian
\begin{eqnarray}\label{v}
H_{\text{eff}1}=\hbar \Omega \left[\sin\left(\frac{\theta}{2}\right)e^{i\varphi} |G\rangle\langle+| - \cos\left(\frac{\theta}{2}\right)|-\rangle\langle+|+ \textrm{H.c.} \right],\notag\\
\end{eqnarray}
with $\theta=2\tan ^{-1}(\Omega_{2}/\Omega_{1})$. Such a $\Lambda$-type energy configuration exhibits the bright and dark states of
$|b\rangle=\sin(\theta/ 2) e^{i\varphi}|G\rangle-\cos(\theta/ 2)|-\rangle$,
$|d\rangle=\cos(\theta/ 2) |G\rangle+\sin(\theta/ 2) e^{-i\varphi}|-\rangle$,
and its dynamics is captured by
\begin{eqnarray}
H_\mathrm{eff1}=\hbar \Omega (|+\rangle\langle b|+ \text{H.c.}),
\end{eqnarray}
that is, a resonant coupling between the bright state $|b\rangle$ and the ancillary state $|+\rangle$ with the dark state $| d \rangle$ being completely decoupled. The evolution operator $U_1$ acting on $|b \rangle$  and $|d \rangle$, thus, results in
\begin{align}
|\psi_{1}(t)\rangle&=U_{1}(t)|d\rangle=|d\rangle, \notag\\
|\psi_{2}(t)\rangle&=U_{1}(t)|b\rangle=\cos(\Omega t)|b\rangle
-i\sin(\Omega t)|+\rangle.
\end{align}
When the condition  $\Omega{\tau_1}=\pi$ is satisfied,  the dressed states undergo a cyclic evolution as
$|\psi_{i}(\tau_1)\rangle\langle\psi_{i}(\tau_1)|
=|\psi_{i}(0)\rangle\langle\psi_{i}(0)|$. Under this condition, the time evolution is given by
\begin{equation}
U_{1}(\tau_1)=\sum\limits^{2}_{i, j=1} \left[ \mathbb{T} e^{i\int_{0}^{\tau_{1}}[A(t)-H_1]dt} \right]_ {i,j}
|\psi_ {i}(0)\rangle \langle\psi_ {j}(0)|,
\end{equation}
where $\mathbb{T}$ is the time-ordering operator and $A_\text{i,j}(t)=i\langle\psi_ {i}(t)|\dot{\psi}_ {j}(t)\rangle$. Meanwhile, as $H_\text{i,j}(t)=\langle\psi_ {i}(t)|H_1|\psi_ t{j}(t)\rangle=0$ is satisfied, there is no transition between the two time-dependent states. Therefore, the induced operation is a  nonadiabatic holonomy matrix
\begin{eqnarray}\label{single}
U_1(\tau_1)=U_1(\theta,\varphi) = \left[
                      \begin{array}{cc}
                        \cos\theta & \sin{\theta}e^{i\varphi} \\
                        \sin{\theta}e^{-i\varphi} & -\cos\theta \\
                      \end{array}
                    \right],
\label{Eqn Usingle}
\end{eqnarray}
in the subspace span $\{|G\rangle, |-\rangle\}$. This gate manifests its geometric feature by its dependence only on the global property of the path but not the traverse detail \cite{SjoqvistNJP2012,XuGF2012PRL}. In addition, as $\theta$ and $\varphi$ can be independently controlled by the two-tone driving $H_{d}$, Eq. (\ref{Eqn Usingle}), thus, pinpoints the arbitrary synthesization of nonadiabatic single-qubit HQC gates.

\begin{figure}[tbp]\centering
\includegraphics[width=8cm]{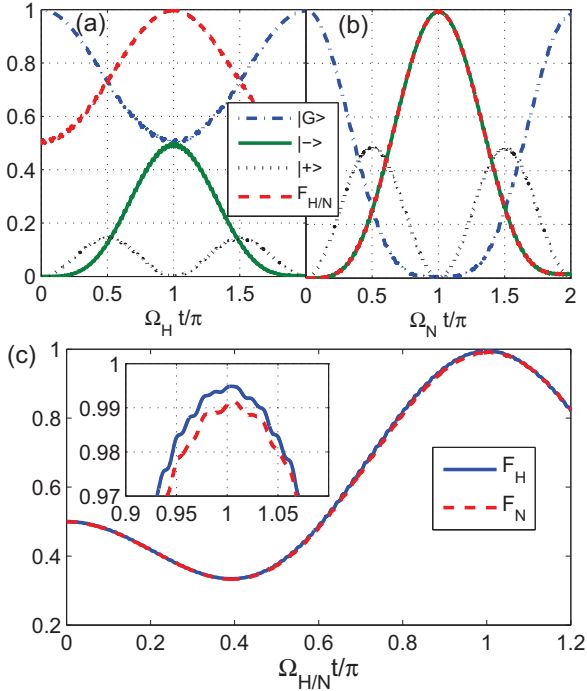}
\caption{State population and fidelity dynamics of the single-qubit operations with the initial state being $|G\rangle$. The results of the Hadamard gate and the NOT gate situations are shown in (a) and (b), respectively. The dynamics of the gate fidelities averaged over $1000$ input states with uniformly distributed $\theta^{\prime}$ is plotted in (c), with the details around the top of the curves shown in the inset.}
\label{Fig f1}
\end{figure}

The performance of the proposed single-qubit gate $U_1(\theta,\phi)$ is mainly limited by the decoherence of the TLR-transmon circuit, the anharmonicity of the transmon, and the leakage of the logic qubit subspace and can be numerically simulated by using the master equation
\begin{eqnarray}  \label{me}
\dot\rho_1 &=& i[\rho_1, H_1] +\frac \kappa  2  \mathcal{L}(a)\notag\\
&&+ \sum_{j=0}^1 \left[ \frac{\Gamma_1^{j}}{2} \mathcal{L}(\sigma^-_{j,j+1}) + \frac{\Gamma_2^{j}}{2}  \mathcal{L}(\sigma^ {z}_{j,j+1})\right],
\end{eqnarray}
where $\rho_1$  is the density matrix of the considered system, $\mathcal{L}(A)=2A\rho_1 A^\dagger-A^\dagger A \rho_1 -\rho_1 A^\dagger A$ is the Lindbladian of the operator $A$, and $\kappa$, $\Gamma_1^j$, $\Gamma_2^j$ denote the decay rate of the TLR, the decay and dephasing rates of the $\{j,j+1\}$ two-level systems, respectively. Because of the finite anharmonicity of the transmon, here we include the third level of the transmon into the numerical simulation by denoting $\sigma^-_{j,j+1}=|j\rangle\langle j+1|$,  $\sigma^{z}_{j,j+1}=|j+1\rangle\langle j+1|-|j\rangle\langle j|$.
Suppose that the qubit is initially prepared in the state $|G\rangle$. We then evaluate the Hadamard and NOT gates using the fidelities defined by $F_H=\langle\psi_f|\rho_1|\psi_f\rangle$ and $F_N=\langle-|\rho_1|-\rangle$, with  $|\psi_f\rangle=(|G\rangle-|-\rangle)/\sqrt{2}$ or $|-\rangle$ being their corresponding target final states. The obtained fidelities are as high as $F_H = 99.71\%$  and $F_N = 99.29\%$  at $t=\pi/\Omega_{H/N}$, as shown in Figs. \ref{Fig f1}(a) and \ref{Fig f1}(b). The parameters of the logic qubit are set as $\omega_c =\omega_q= 2\pi \times 6$ GHz, $g_0/2\pi=300$ MHz, $\Omega_H = \Omega_N=2\pi\times 8$ MHz,  and $\Gamma_1^j=\Gamma_2^j= \kappa=2\pi \times 10$ kHz corresponding to the coherent time of 16 $\mathrm{\mu s}$ accessible with the current level of technology \cite{DevoretScienceReview2013,mjp}. The anharmonicity of the third level is set as $2\pi\times 310$ MHz \cite{AbdumalikovNature2013}. For the  Hadamard gate, we modulate $\Omega_1/ \Omega_H \simeq 0.924$ and $\Omega_2/ \Omega_H \simeq 0.383$ to ensure $\theta = \pi/4$, while for the NOT gate we choose $\Omega_1=\Omega_2= \Omega_N/\sqrt{2}$. Our numerical results indicate that the infidelity is mainly due to the decoherence, the limitation on the anharmonicity of transmons, and the leakage of the logical qubit subspace.

It should be emphasized that our numerical calculation is based on the full Hamiltonian $H_1$ in Eq. (\ref{h1}) and does not rely on any further approximation. Moreover, the interactions between the higher levels of the transmon and the TLR mode and the effects of the two-tone driving in the expanded Hilbert subspace are taken into account. In addition, for a general initial state $|\psi\rangle=\cos\theta'|G\rangle+\sin\theta'|-\rangle$ with $\theta'=0$ corresponding to the ground state, we numerically confirm that the fidelity depends weakly on $\theta'$. To fully quantify the performance of the implemented gate, we plot in Fig. \ref{Fig f1}(c) the gate fidelities for $1000$ input states with $\theta'$ uniformly distributed over $[0, 2\pi]$,
where we find that $F_H ^G=99.49\%$ and  $F_N ^G=99.15\%$ which are higher than the threshold of surface code error correction schemes.

\section{The nontrivial two-qubit gate}

We next consider the implementation of two-qubit HQC gates between the neighboring logic qubits 1 and 2 in Fig. \ref{Fig setup}(a). This can be achieved by the ancillary of the third logic qubit 3, which shares the same grounding SQUID with the two target qubits. Without loss of generality, here we set  the TLR-transmon coupling $g_1=g_2=g_3 =g=2\pi \times 100$ MHz among the three logic qubits. When the grounding SQUID is dc biased,
the linear coupling between the three TLRs can be reduced to
\begin{eqnarray}  \label{Eqn coupled}
H_\mathrm{dc}&=&  \hbar \left( \mathcal{J}_{12} a_1^\dagger a_2 +\mathcal{J}_{23} a_2^\dagger a_3+\mathcal{J}_{31} a_3^\dagger a_1\right) +\mathrm{H.c.}\notag\\
&=& \frac{\hbar}{2} \sum_{j} \mathcal{J}_{j,j+1} \left(|G-\rangle+|G+\rangle \right)_{j,j+1}\notag\\
&&\times  \left(\langle-G| + \langle+G|\right)+\mathrm{H.c.},
\end{eqnarray}
in the dressed-states subspace, with $\mathcal{J}_{j,j+1} \ll \delta_c$ being the dc coupling strength induced by the grounding SQUID (see Appendix A for details). Because of the large detuning $\delta_c$, the static exchange coupling $H_\mathrm{dc}$ does not produce a significant effect. Meanwhile, we can exploit the alternative dynamic modulation method \cite{FanSHDMNP2012,NISTParametricConversionNP2011,NISTHongOuMandelPRL2012, NISTHongOuMandelPRL2012,DCEexperimentNature2011}: The grounding SQUID can be regarded as a tunable inductance which can be ac modulated by external magnetic flux oscillating at very high frequency \cite{DCEexperimentNature2011}. Such ac modulation introduces a small fraction
\begin{equation} \label{hac}
 H_\mathrm{ac}=\sum_{j} \hbar \mathcal{J}^{\mathrm{ac}}_{j,j+1}(t)(a_j^\dagger a_{j+1}+\mathrm{H.c.}),
\end{equation}
in addition to the irrelevant dc $H_\mathrm{dc}$ (see Appendix B for details). The modulating frequency of $\Phi^{\mathrm{ac}}_{\mathrm{ex}}(t)$ must be lower than the plasma frequency $\omega_{p}$ of the grounding SQUID \cite{KochTransmonPRA2007}, otherwise the internal degrees of freedom of the SQUID will be activated and complex quasiparticle excitations will emerge \cite{FelicettiPRL2014}. In our setup, the condition $\omega_{p}\gg\delta_c$ is well fulfilled, and the excitation of the grounding SQUID is highly suppressed.

Generally, we may assume that the ac modulation of the grounding SQUID contains two tones which induce the excitation exchange of $|-G\rangle _{1,3}\leftrightarrow|G+\rangle _{1,3}$ and $|-G\rangle _{2,3}\leftrightarrow|G+\rangle _{2,3}$ by bridging their frequency gaps, respectively. However, with our prescribed TLR frequencies and identical TLR-transmon coupling strength, the two target transitions are of the same frequency gap (see Appendix C for details), and, thus, they can be induced by a single frequency ac modulation. In the rotating frame of $H_\mathrm{JC}$, $H_\mathrm{ac}$ can then be reduced to
\begin{eqnarray}  \label{h2}
H_\mathrm{2}= \hbar \mathcal{T} (|- G\rangle _{1,3}\langle G+|
 +|- G\rangle _{2,3}\langle G+| ) +\mathrm{H.c.},
\end{eqnarray}
where $\mathcal{T}/2\pi \in  \left[5, 10\right] \textrm{MHz}$ is the parametric coupling strength induced by the parametric modulation. The other allowed transitions in $H_\mathrm{ac}$ are detuned at least by $2g$ and can thus be safely neglected by the rotating-wave approximation. Similar to the single-qubit case, we can figure out that the single excitation subspace span$\{|-GG\rangle_{1,2,3}, |G-G\rangle_{1,2,3}, |GG+\rangle_{1,2,3}\}$ constitutes a three-level system. When the cyclic condition $\int_0^{\tau} J dt=\pi$ with $J=\sqrt{2} \mathcal{T}$ is fulfilled, a holonomic quantum gate
\begin{eqnarray}  \label{u2}
U_2=\left(
\begin{array}{cccc}
         1 & 0 & 0 & 0 \\
        0 & 0 & 1 & 0 \\
            0 & 1 & 0 & 0 \\
            0 & 0 & 0 & -1\\
          \end{array}
\right),
\end{eqnarray}
can be induced in the Hilbert subspace  span$\{|GG\rangle_{1,2}, |G-\rangle_{1,2}, |-G\rangle_{1,2}, |--\rangle_{1,2}\}$. The combination of $U_2$ and $U_1(\theta, \varphi)$, thus, forms a universal set of quantum gates. We note that the minus sign for the element $|--\rangle_{1,2}\langle--|$ in Eq. (\ref{u2}) comes from the holonomic dynamics of another subspace  span$\{ |--G\rangle_{1,2,3}, |-G+\rangle_{1,2,3}, |G-+\rangle_{1,2,3} \}$, which has the same energy spectrum as that of the two-qubit gate subspace  span$\{|-GG\rangle_{1,2,3}, |G-G\rangle_{1,2,3}, |GG+\rangle_{1,2,3}\}$. Within this subspace,  the $|--\rangle_{1,2}$ state obtains a $\pi$ phase during the implementation of the two-qubit gate in Eq. (\ref{u2}).

\begin{figure}[tbp] \centering
\includegraphics[width=7cm]{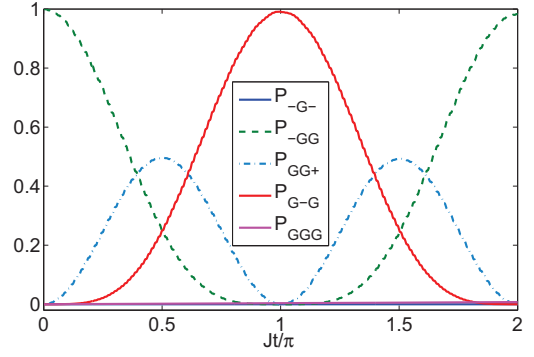}
\caption{State population and fidelity dynamics of the $U_2$ gate as a function of $J t/\pi$ with the initial state being $|-GG\rangle_{1,2,3}$.}
\label{twobit}
\end{figure}

Similarly, we further verify the performance of the two-qubit gates by taking  $\mathcal{T}/2\pi= 6$ MHz. We calculate the state populations and fidelity for an initial state $|-GG\rangle_{1,2,3}$ using the Hamiltonian $H_{\mathrm{ac}}$ in Eq. (\ref{hac}) and plot the fidelity dynamics of $F_T=_{1,2,3}\langle G-G|\rho_2|G-G\rangle_{1,2,3}$ with $\rho_2$ being the time-dependent density matrix of the considered two-qubit system. As shown in Fig. \ref{twobit}, the obtained fidelity is comparable to that of the single-qubit operations, with a fidelity as high as $F_T=99.09\%$. This fidelity is in sharp contrast with the existing implementations and can be interpreted in an intuitive way: As the interactions exploited in our scheme are resonant, the speed of two-qubit gate is comparable to the case of the single-qubit gate, which is distinct from the previous schemes using dispersive couplings.

\section{Discussion}
Our scheme can be readily scaled up to facilitate the scalability criteria of quantum computing. As shown in Fig. \ref{Fig setup}(a), we can form a 2D array of the logic qubit by placing the TLRs in an interlaced honeycomb lattice. This configuration allows the holonomic two-qubit gates  to be established between any two logic qubits sharing the same grounding SQUID with the third one serving as ancillary. With regard to the feasibility of current proposal, we first notice that the elementary gates involve the control of both the SQUIDs of the transmon qubit and the grounding SQUIDs. This is well within the reach of the current level of technology, as both the dc and ac flux controls have already been achieved in coupled superconducting qubits with both the loop sizes and their distances being at the range of micrometers  \cite{PlantenbergCNOTNature2006,PloegCoupleFQ2007}. As for the scaled lattice, the individual control, wiring, and readout can be achieved by adding an extra layer on the top of the qubit lattice layer \cite{barends2014,YaleMultilayerNPJQI2016,YaleMultilayerPRA2016}, and the interlayer connection can be obtained by the capacitive coupling. In addition, the parametric coupling exploited in our scheme has been demonstrated previously in few-body systems \cite{NISTParametricConversionNP2011,NISTHongOuMandelPRL2012,NISTParametricCouplingPRL2014,NISTCoherentStateAPL2015} and recently in a SQC lattice, with a synthetic gauge field for the microwave photons being observed \cite{RoushanChiral2016}. This experimental progress, thus, partially verifies the feasibility of our scheme. Finally, the fluctuation induced by the ubiquitous flicker noises in the SQC should also be considered \cite{FlickerRMP2014}. We notice that the proposed circuit is insensitive to the charge noise as it consists of only linear TLRs, grounding SQUIDs with very small anharmonicity and the charge-insensitive transmon qubits \cite{KochTransmonPRA2007}. For the flux-type and critical current-type $1/f$ noise, their influence is estimated to be much weaker than the decay effect \cite{WangYPChiral2015,WangYPNPJQI2015,WangYPLieb2016}, which has already been included in our numerical simulations.

\section{Conclusion}
In summary, we propose a scheme of quantum computation with dressed-state qubits in circuit QED using nonadiabatic holonomies. In particular, the single-qubit gates can be achieved through external microwave-driving fields and the two-qubit gates can be obtained in a fast resonant way. Therefore, our scheme presents a promising way of realizing robust and efficient HQC in superconducting devices.


\acknowledgements
This work is supported by the NFRPC (Grant No. 2013CB921804), the NKRDPC (Grants No. 2016YFA0301200 and No. 2016YFA0301803), the NSFC (Grants No. 11104096, No. 11374117 and No. 11604103), NSAF (Grants No. U1330201, and No. U1530401), and the NSF of Guangdong Province (Grant No. 2016A030313436).

\appendix

\section{The dc mixing induced by the grounding SQUID}
In this appendix, we derive in detail the coupling between the logic qubits through the detailed analysis of a three-qubit unit cell of the proposed circuit lattice. During this investigation, we also estimate the parameters of the proposed circuit based on recently reported experimental data \cite{DCEexperimentNature2011,NISTParametricConversionNP2011,NISTHongOuMandelPRL2012,
NISTParametricCouplingPRL2014,NISTCoherentStateAPL2015} and propose their representative values, as listed in Table \ref{Tab para}. The influence from the other part of the lattice is temporarily minimized by setting the grounding SQUIDs at the individual ends of the three TLRs with infinitesimal effective inductances.

We assume the common grounding SQUID of the unit cell has an effective Josephson energy of $E_{J}=E_{J0}\cos( \pi \Phi_{\mathrm{ext}} / \phi_{0})$ with $E_{J0}$ being its maximal Josephson energy, $\Phi_{\mathrm{ext}}$ the external flux bias, and $\phi_{\mathrm{0}}=h/2e$ the flux quanta. In the first step, let us assume that only a dc flux bias $\Phi_{\mathrm{ex}}^\mathrm{dc}$ is added. Physically, a certain TLR  can hardly "feel" the other two  TLRs as the currents from them will flow mostly to the ground through the SQUID due to its very small inductance \cite{FelicettiPRL2014,WangYPChiral2015}. The SQUID can then be regarded as a low-voltage shortcut of the three TLRs  and, thus, allows the definition of individual TLR modes in this  unit cell; see Refs. \cite{xue2,WangYPLieb2016} for details. Meanwhile, the eigenmodes are well separated in the corresponding TLRs, indicating the one-to-one correspondence between the TLRs and the eigenmodes. Furthermore, these eigenmodes can well be approximated by the $\lambda/2$ mode of the TLRs with the nodes located at the nodes, which is consistent with the described shortcut boundary condition. In addition, the eigenmodes  can be quantized as
\begin{align}
\mathcal{H}_{\mathrm{0}}=\sum _{m}\hbar \omega _{cm}(a_{m}^{\dag }a_{m}+\frac{1}{2}),
\label{eq:eigenhamiltonian}
\end{align}
where $\omega _{cm}$, $a_{m}^{\dag}$, and $a_{m}$ are the frequency, creation, and annihilation operators of the $m$th eigenmode.

\begin{table}[tb]
\centering
\caption{Representative parameters of the proposed circuit selected based on recently-reported experiments.}
\begin{tabular}{p{0.22\textwidth}p{0.25\textwidth}}
  \hline
  TLRs parameters &  \\
  \hline
  unit inductance & $l=4.1\times 10^{-7}\,\mathrm{H}\cdot \mathrm{m}^{-1}$ \cite{NISTParametricConversionNP2011,NISTHongOuMandelPRL2012,NISTParametricCouplingPRL2014}\\
  unit capacitance & $c=1.6\times 10^{-10}\,\mathrm{F}\cdot \mathrm{m}^{-1}$ \cite{NISTParametricConversionNP2011,NISTHongOuMandelPRL2012,NISTParametricCouplingPRL2014}\\
  lengths  &  $L_{1}=10.2\,\mathrm{mm}$\\
  & $L_2=8.5\,\mathrm{mm}$ \\
                      &    $L_3=9.57\,\mathrm{mm}$
  \cite{DCEexperimentNature2011,NISTParametricConversionNP2011,NISTHongOuMandelPRL2012}
  \\
  \hline
  SQUIDs &\\
  \hline
  maximal critical currents & $I_{\mathrm{J0}}=46\,\mu \mathrm{A}$ \cite{DCEexperimentNature2011,NISTParametricConversionNP2011, YuYangScience2002,MartinisPhaseQubitPRL2002}\\
  dc flux bias points & $\Phi_\mathrm{ex}^\mathrm{dc}=0.43\Phi_0$ \cite{NISTParametricConversionNP2011,NISTHongOuMandelPRL2012}\\
  effective critical currents & $I_{\mathrm{J}}=10\,\mu \mathrm{A}$\\
  junction capacitances & $C_\mathrm{J}=0.5$ $\mathrm{pF}$ \cite{YuYangScience2002,MartinisPhaseQubitPRL2002}\\
  ac modulation amplitudes & $\Phi_{13}=1.53\%\Phi_0$  \\
                    &  $\Phi_{23}=1.66\%\Phi_0$ \cite{NISTParametricConversionNP2011}\\
  \hline
  Eigenmodes and coupling & \\
  \hline
  eigenfrequencies & $\omega_{c1}/2\pi=6 \,\mathrm{GHz}$ \\
  & $\omega_{c2}/2\pi=7.2$ GHz \\
  & $\omega_{c3}/2\pi=6.4$ GHz \cite{DCEexperimentNature2011,NISTParametricConversionNP2011,NISTHongOuMandelPRL2012}\\
  uniform decay rate & $\kappa/2\pi=10\,\mathrm{kHz}$ \cite{DevoretScienceReview2013}\\
  hopping strengths & $\mathcal{T}_{13}/2\pi=\mathcal{T}_{23}/2\pi=6$ $\mathrm{MHz}$\\
  \hline
\end{tabular}
\label{Tab para}
\end{table}

Here, we temporarily stop to check the role played by the grounding SQUID. First, the gauge-invariant phase difference of the SQUID can be written as
\begin{equation}
\label{eq:phiJ1}
\phi_{J}=\sum_{m} \phi^{m}(a_m+a_m^\dagger),
\end{equation}
where $\phi^{m}=f_{\alpha,m}(x=L_\alpha)\sqrt{\hbar/2\omega_{cm} c}$ is the rms node flux fluctuation of the $m$th mode across the SQUID with
\begin{align}
(\phi^1,\phi^2,\phi^3)/\phi_0 =(3.6,3.4,3.1)\times 10^{-3}.
\end{align}
Such small fluctuation of $\phi_{J}$ verifies the linearized treatment of  the grounding SQUID in the quantization of the eigenmodes and indicates that the eigenmodes can be regarded as the individual $\lambda/2$ modes of the TLRs slightly mixed by the grounding SQUID with small inductance.

We then proceed to estimate to what extent the grounding SQUID mixes the individual $\lambda/2$ modes of the TLRs, which is due to the dc Josephson coupling
\begin{align}
\mathcal{E}_{\mathrm{dc}}&=-E_{{J}}\cos \left( \frac{\phi_{J}}{\phi_{0}} \right)
\approx \frac{1}{2} \left( \frac{\phi_{J}}{\phi_{0}} \right)^2 E_{J0}
\cos \left(\frac{\Phi_{\mathrm{ex}}^{\mathrm{dc}}}{2\phi _{0}}\right)
\notag\\
&=\hbar \sum _{m,n}\mathcal{J}_{mn}^{\mathrm{dc}}(a_{m}^{\dagger }+a_{m})(a_{n}^{\dagger }+a_{n}),
\label{eq:EJosillation}
\end{align}
with $\Phi_{\mathrm{ex}}^{\mathrm{dc}}$ being the external dc magnetic flux, and the coupling strength between two eigenmodes is
\begin{eqnarray}
\mathcal{J}_{mn}^{\mathrm{dc}}=\frac{\phi^m\phi^n}{\phi_0^2}E_{J0}\cos \left(\frac{\Phi_{\mathrm{ex}}^{\mathrm{dc}}}{2\phi _{0}}\right).
\end{eqnarray}
$\mathcal{J}_{mn}^{\mathrm{dc}}$ can then be regarded as the dc mixing between the individual $\lambda/2$ modes induced by the static bias of the grounding SQUID.
As
\begin{equation}
\mathcal{J}_{mn}^{\mathrm{dc}} \simeq  2\pi \times 56 \text{ MHz} < \delta_{c}/7,
\end{equation}
the grounding SQUID can slightly mix  the original modes of the TLRs.

We can also estimate the higher fourth-order nonlinear term of $-E_{J0}\cos(\phi_{J}/\phi_0)$ as
\begin{align}
    \mathcal{E}_{\mathrm{dc}}^4 \approx \frac{1}{48} \left( \frac{\phi^j}{\phi_{\mathrm{0}}} \right)^4 E_{J0}\cos \left(\frac{\Phi_{\mathrm{ex}}^{\mathrm{dc}}}{2\phi _{0}}\right)
     \sim 10^{-6} \mathcal{J}^{\mathrm{dc}}_{mn},
\end{align}
i.e., 6 orders of magnitude smaller than the second-order terms reserved in Eq. (\ref{eq:EJosillation}) and, thus, verifies the validity of keeping only the second-order terms in Eq.~(\ref{eq:EJosillation}).

In addition, we can observe that $\mathcal{J}_{mn}$ scales versus $E_{J0}$ as $\mathcal{J}_{mn}^{\mathrm{dc}} \propto E_{J0}^{-1}$ with increasing $E_{J0}$. This can be interpreted by the role of the grounding SQUID. Because of the low-inductance shortcut boundary condition, the node flux $\phi_\mathrm{J}$ across the grounding SQUID scales as $E^{-1}_\mathrm{J0}$; thus, the coupling energy $E_{J0}\cos(\phi_{J}/\phi_{0})\approx -\phi_{J}^2/2L_{J} \propto E_{J0}^{-1}$. This scaling behavior provides an efficient way of suppressing the unwanted cross talk on the lattice:  One can isolate a part of the lattice (e.g., a few logic qubits) by simply tuning up the Josephson energies of the grounding SQUIDs it shares with the other parts.

\section{Parametric coupling between the eigenmodes}

The parametric coupling between the three logic qubits originates from the dependence of $E_{\mathrm{J}}$ on the total external magnetic flux $\Phi_{\mathrm{ext}}= \Phi_{\mathrm{ex}}^{\mathrm{dc}}
+\Phi_{\mathrm{ex}}^{\mathrm{ac}}(t)$,
\begin{align}
E_{J}&=E_{{J0}}\cos \left(\frac{\Phi_{\mathrm{ext}}}{2\phi _{0}}\right) \notag \\
&\approx E_{J0}\cos \left(\frac{\Phi_{\mathrm{ex}}^{\mathrm{dc}}}{2\phi _{0}}\right)-\frac{E_{J0}\Phi_{\mathrm{ex}}^{\mathrm{ac}}(t)}{2\phi _{0}}\sin \left(\frac{\Phi_{\mathrm{ex}}^{\mathrm{dc}}}{2\phi _{0}}\right),
\label{eq:EJosillation2}
\end{align}
where we assume that a small ac fraction $\Phi_{\mathrm{ex}}^{\mathrm{ac}}(t)$ is added to $\Phi_{\mathrm{ext}}$ with $\left|\Phi_{\mathrm{ex}}^{\mathrm{ac}}(t)\right| \ll \left|\Phi_{\mathrm{ex}}^{\mathrm{dc}} \right|$. We first consider the case of omitting the transmons (e.g., by tuning them far off resonant with their TLRs) and assume that  $\Phi_{\mathrm{ex}}^{\mathrm{ac}}(t)$ is composed of two tones \begin{align}
\label{Eqn twotone}
\Phi_{\mathrm{ex}}^{\mathrm{ac}}(t)&=\Phi_{13}\cos ( \omega_1 t )+\Phi_{23}\cos ( \omega_2 t  ),
\end{align}
where the $\omega_1$ tone is exploited to induce the $ 1 \Leftrightarrow 3 $ hopping, and the $\omega_2$ tone is used for the $2 \Leftrightarrow 3$ hopping. By representing $\phi_{J}$ as the form shown in Eq.~(\ref{eq:phiJ1}), we obtain the ac coupling from the second term of Eq.~ (\ref{eq:EJosillation2})
\begin{equation}
H_\mathrm{ac}=\frac{ E_\mathrm{J0} \Phi_{\mathrm{ex}}^{\mathrm{ac}}(t)} {4\phi _{0}^3} \sin \left(\frac{\Phi_{\mathrm{ex}}^{\mathrm{dc}}}{2\phi _{0}}\right)
\left[\sum _{m} \phi^m \left(a_m+a_m^{\dagger}\right)\right]^2.
\label{eq:HDCcoupling2}
\end{equation}
In the rotating frame of $H_0$, the induced parametric photon hopping between the TLRs can be further written as
\begin{align}
\label{Eqn effHami}
H_{\mathrm{eff2}}&=e^{itH_0} {H}_{\mathrm{ac}} e^{-itH_0} \notag\\
& \simeq 2\hbar \left(\mathcal{T}_{1,3}a_1^{\dagger}a_3
+\mathcal{T}_{2,3}a_2^{\dagger}a_3 \right)+\mathrm{H.c.},
\end{align}
where $2\mathcal{T}_{m,n}$ are the effective hopping strengths proportional to the corresponding $\Phi_{mn}$ in Eq. (\ref{Eqn twotone}), and other fast-oscillating terms can be omitted due to the rotating-wave approximation. The amplitudes of the two tones can be selected in the range $\left[ \Phi_{\mathrm{13}}, \Phi_{\mathrm{23}}\right]=\Phi_0\left[ 1.53\%, 1.66\% \right]$ such that the coupling strength $\mathcal{T}_{m,n}/2\pi \in [5, 10]\,\mathrm{MHz}$ can be induced   \cite{NISTParametricConversionNP2011,NISTHongOuMandelPRL2012,NISTParametricCouplingPRL2014, NISTCoherentStateAPL2015}.

\begin{figure}[tbp] \centering
\includegraphics[width=6.5cm]{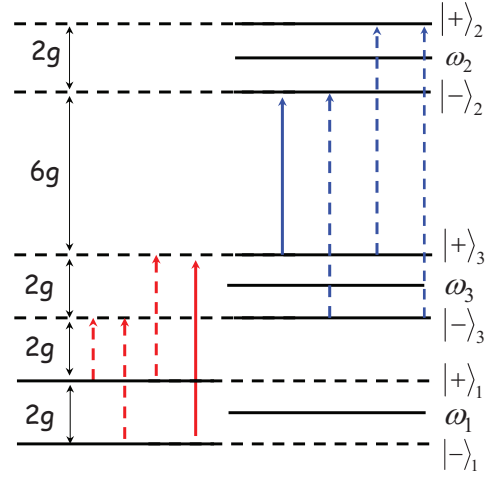}
\caption{The allowed transitions of the three coupled TLRs system with the two target transitions are indicated by solid arrows.}
\label{two}
\end{figure}

\section{The two-qubit gates}

The described parametric coupling scheme is not influenced much by the inclusion of the transmons. We recall that the dressed states $|-\rangle$ of the logical qubits are half-TLR plus half-transmon excitation; therefore, the parametric hopping of these states can be directly induced by the parametric coupling of their photonic component. In this situation, we just need to adjust the two-tone pulse to fill the gaps between the transitions of $1 \Leftrightarrow 3 $ and $ 2 \Leftrightarrow 3$ and enlarge the amplitudes of the tones by twice, as the dressed states contain only half-TLR components. Explicitly, when transmons are loaded into each of the TLRs, the energy spectrum splits. However, the parametric coupling can still induce relevant transitions. We now present an example with two TLRs. We still set the parameters of the first TLR-transmon unit as  $\omega_{c,1}=2\pi\times 6$ GHz and $g_1=g=2\pi\times 100$ MHz. The third ancillary TLR-transmon unit is designed to be $\omega_{c,3}=2\pi\times 6.4$ GHz and $g_3=g$. By these settings, the energy spectrum of the two-cavity system is shown in Fig. \ref{two}. Similar to the discussion above, the two TLRs are coupled in an exchanged manner as \begin{eqnarray}  \label{twocoupled}
H_{\text{1,3}}&=& \hbar \mathcal{J}_{13}^{\text{ac}}(t) a_1^\dagger a_3 + \text{H.c.}\\
&\equiv& {\hbar\mathcal{J}_{13}^{\text{ac}}(t) \over 2}   \left(|G-\rangle+|G+\rangle \right)_{1,3}  \left(\langle-G| + \langle+G|\right)+\text{H.c.},\notag
\end{eqnarray}
which means that the four transitions indicated by red lines,  both solid and dashed, are allowed. However, as $\mathcal{J}_{13}^{\mathrm{dc}}\ll g$, direct transition is not allowed due to the existence of the energy mismatch. To see this, we transform $H_\text{1,3}$ in Eq. ({\ref{twocoupled}}) into the interaction picture with respect to
\begin{eqnarray}  \label{free}
H_{\text{0}}= \hbar \sum_{j=1}^2  \left( \omega_{-,j} |-\rangle_j\langle-| +\omega_{+,j} |+\rangle_j\langle+| \right).
\end{eqnarray}
The transformed Hamiltonian is
\begin{eqnarray}
H_{\text{1,3}}&=& {\hbar \mathcal{J}_{13}^{\text{ac}}(t)  \over 2} (|G-\rangle_{1,3}\langle+G|e^{2 igt} +|G-\rangle_{1,3}\langle-G|e^{4igt}\notag\\
&+& |G+\rangle_{1,3}\langle+G|e^{4igt}
+|G+\rangle_{1,3}\langle-G|e^{6igt} )+\text{H.c.}. \notag
\end{eqnarray}
To induce the transition of $|-G\rangle_{1,3}\leftrightarrow|G+\rangle _{1,3}$, we set $\mathcal{J}_{13}^{\text{ac}}(t)=4\mathcal{T}_{13}\cos(6gt)$. In this case, other allowed transitions will be detuned at least by $2g$.

For the two-qubit-gate purpose, we set the parameters of the second TLR-transmon unit to be $\omega_{c,2}=2\pi\times 7.2$ GHz, $g_2=g$, and $\mathcal{J}_{23}^{\text{ac}}(t)=4\mathcal{T}_{23}\cos(6gt)$, which lead to the transition of $|-G\rangle_{2,3}\leftrightarrow|G+\rangle _{2,3}$. Therefore,  we need only to ac modulate the grounding SQUID with a single frequency, i.e.,
$\mathcal{J}_{\text{ac}}(t)=4\mathcal{T}\cos(6gt)$.

\end{document}